\documentclass[11pt,twoside]{article}
\usepackage{asp2010}

\resetcounters

\begin{document}

\title{Fundamental Stellar Properties from Optical Interferometry}
\author{Gerard~T.~van~Belle$^1$, Jason~Aufdenberg$^2$, Tabetha~Boyajian$^{3,9}$, Graham~Harper$^4$, Christian~Hummel$^1$, Ettore~Pedretti$^5$, Ellyn~Baines$^6$, Russel~White$^3$, Vikram~Ravi$^7$, Steve~Ridgway$^8$
\affil{$^1$European Southern Observatory, Karl-Schwarzschild-Str. 2, 85748 Garching, Germany}
\affil{$^2$Embry-Riddle Aeronautical University, 600 S. Clyde Morris Blvd. Daytona Beach, Florida 32114, USA}
\affil{$^3$Department of Physics and Astronomy,
      Georgia State University,
      P.O. Box 4106,
      Atlanta, GA 30302-4106, USA}
\affil{$^4$Astrophysics Research Group,
School of Physics, Trinity College, Dublin 2, Ireland}
\affil{$^5$School of Physics and Astronomy,
University of St Andrews,
North Haugh,
St Andrews,
Fife,
KY16 9SS, UK}
\affil{$^6$Remote Sensing Division, Naval Research Laboratory, 4555 Overlook Avenue SW, Washington, DC 20375, USA}
\affil{$^7$University of California, Berkeley
Space Sciences Lab
7 Gauss Way
Berkeley, CA 94720-7450, USA}
\affil{$^8$Kitt Peak National Observatory
National Optical Astronomy Observatories
P.O. Box 26732
Tucson, AZ 85726-6732, USA}
\affil{$^9$Hubble Fellow}
}

\begin{abstract}
High-resolution observations by visible and near-infrared interferometers of both single stars and binaries have made significant contributions to the foundations that underpin many aspects of our knowledge of stellar structure and evolution for cool stars.  The CS16 splinter on this topic reviewed contributions of optical interferometry to date, examined highlights of current research, and identified areas for contributions with new observational constraints in the near future.
\end{abstract}

\section{Introduction}

Observations of the fundamental parameters of stars - their masses, radii and effective temperatures - are crucial components in understanding stellar formation and evolution.  For many types of stars, obtaining measurements with sufficient precision to constrain theoretical models can be challenging.  A discussion of the contributions of optical interferometry to this field is timely given the technological maturation and recent scientific contributions of modern facilities such as VLTI, the Keck Interferometer, and the CHARA Array.  Furthermore, a promising road lies ahead, as the 2-telescope measurements used in most of the work described below, are now being expanded by multi-way multi-channel combiners at the VLTI and CHARA which provide significant additional constraints by adding closure phase, differential phase, and in a few cases, imaging.

Areas where optical interferometry is making compelling contributions towards parameterizing the fundamental properties of cool stars include:
{\bf Radius} - The sub-milliarcsecond resolutions of modern optical interferometers allow for direct measurement of the linear sizes of nearby cool stars of all luminosity classes.
{\bf Effective temperature} - Angular sizes, in combination with measured spectral energy distributions, provide direct quantification of this macroscopic quantity.
{\bf Mass} - Dynamical masses from orbits determined with optical interferometers and radial velocity measurements are the highest-precision determinations of this most important fundamental parameter. Additionally, asteroseismological measurements of single stars provide direct measures of stellar mean density; coupled with interferometric radii, mass determinations are possible.
{\bf Distance} - Parallactic measurements of stellar distances from optical interferometers calibrate stellar luminosities.
{\bf Temperature Structure} - Limb-darkening measurements from optical interferometers probe the vertical temperature structure of a stellar atmosphere, providing constraints on model atmosphere structures in general, and models for convective flux transport in particular.


Specific scientific questions that were considered during the splinter session were:
(1) What have been the constraints on fundamental parameters for cool stars provided by optical interferometry?  What are the limits on those constraints?
(2) Where are the most attractive areas for guiding development of cool stars astrophysical models with observations from optical interferometers?
(3) Are there specific areas of cool star evolution that are particularly well suited for studies by optical interferometry?
(4) What are the observing opportunities available today for cool star research?
(5) What are the prospects for future developments in optical interferometry that can significantly advance our knowledge of cool stars?


Answers to these questions are of considerable interest to the Cool Stars 16 audience.  Furthermore, the most recent event formally addressing fundamental stellar parameters was over a decade ago\footnote{IAU 189 in Sydney, Australia, ``Fundamental Stellar Properties: the Interaction Between Observation and Theory'' \citep{Bedding1997IAUS..189.....B}}, predating the current generation of operational facilities, and did not specifically address cool stars.

\section{Angular Sizes}

With available mass estimates, temperature and size predictions of stellar evolutionary models as a function of age can be compared to values obtained with interferometry.  Historically, this had been limited to eclipsing binary systems \citep{Andersen1991A&ARv...3...91A}.  Long-baseline optical/infrared interferometry has changed this dramatically by providing sufficient resolution to measure the angular sizes of dozens of nearby stars.  Angular size measurements with errors under 1\% are now possible.  The limb-darkening corrections, normally modeled, are now subject to direct interferometric verification \citep[e.g.][]{Lacour2008A&A...485..561L}.

Measurements of single M-dwarf radii show that they are potentially 10-15\% larger than currently predicted by models
\citep{Berger2008ASPC..384..226B} with a suggestion that the discrepancy increases with elevated metallicity, although not all studies are in agreement on this point \citep{Demory2009A&A...505..205D}.  Examples of ultra-precise radii and temperatures have been measured for coeval binary stars \citep{Kervella2008A&A...488..667K}, metal poor population II stars \citep{Boyajian2008ApJ...683..424B}, giant stars in the Hyades \citep{Boyajian2009ApJ...691.1243B}, pulsating Mira variables \citep{Thompson2002ApJ...577..447T,vanBelle2002AJ....124.1706V}, exoplanet host stars \citep{vanBelle2009ApJ...694.1085V} and calibration of the giant and supergiant effective temperature scales \citep[][respectively]{vanBelle1999AJ....117..521V, vanBelle2009MNRAS.394.1925V}.   In the precise measurements of radii, interferometry has also revealed very low levels of circumstellar emission around main sequence stars
\citep{Ciardi2001ApJ...559.1147C,Absil2008A&A...487.1041A,Akeson2009ApJ...691.1896A} that are too faint and close to the star to be otherwise detected.

Furthermore, interferometry is showing how limited the very concept of effective temperature can be.  Rapidly rotating stars have significant temperature gradients on their surfaces, but even more leisurely stars like Procyon cannot be fit precisely with a single $T_{\rm EFF}$ model.  For Procyon, multi-wavelength angular diameters from Mark III and VLTI showed that 1-D mixing length convection doesn't correctly predict stellar limb darkening (Aufdenberg et al. 2005).  Recent 3-D models of the solar atmosphere \citep{Asplund2009ARA&A..47..481A} have significantly revised the solar abundance;  such 3-D models for more distant stars can be tested with interferometric limb darkening measurements in ways that spectroscopy alone cannot.

\subsection{Observational Results of Diameters, Temperatures - Tabetha Boyajian}

Studies targeting the discovery of previously unknown nearby stars, such as RECONS \citep{Henry2006AJ....132.2360H,Subasavage2008AJ....136..899S}, have uncovered a wealth of new objects to be examined with optical interferometers - particularly cool main sequence stars.  Angular diameters measured by interferometry can be combined with distances (well known for nearby objects) to produce linear radii; with bolometric fluxes to directly produce effective temperatures ($T_{\rm EFF} \propto (F_{\rm BOL} / \theta^2)^{1/4}$).
There are well over 500 objects for which angular sizes have been measured \citep{Richichi2005A&A...431..773R}, roughly a four-fold increase over just a decade ago \citep{Davis1997IAUS..189...31D}, with many of the new measurements being produced for main sequence and not evolved objects \citep{Boyajian2009PhDT........24B}.

For stars similar to the Sun, comparison of the stellar effective temperature measured directly versus semi-empirically \citep{AllendePrieto1999A&A...352..555A,Holmberg2007A&A...475..519H,Takeda2007PASJ...59..335T}
show the temperature values are lower (by $\sim 1.5-4$\%), and radii values are higher (by $\sim 4 - 10$\%) then those semi-empirical approaches \citep{Boyajian2009PhDT........24B}.  These offsets are such that the resulting luminosities, regardless of method, are in agreement with each other.  Additionally, \citet{Boyajian2009PhDT........24B} point out that no correlation is seen when investigating the deviation in temperature and radii with respect to color index or metallicity.

\begin{figure}[!ht]
\label{fig:M_VS_L}
\plotone{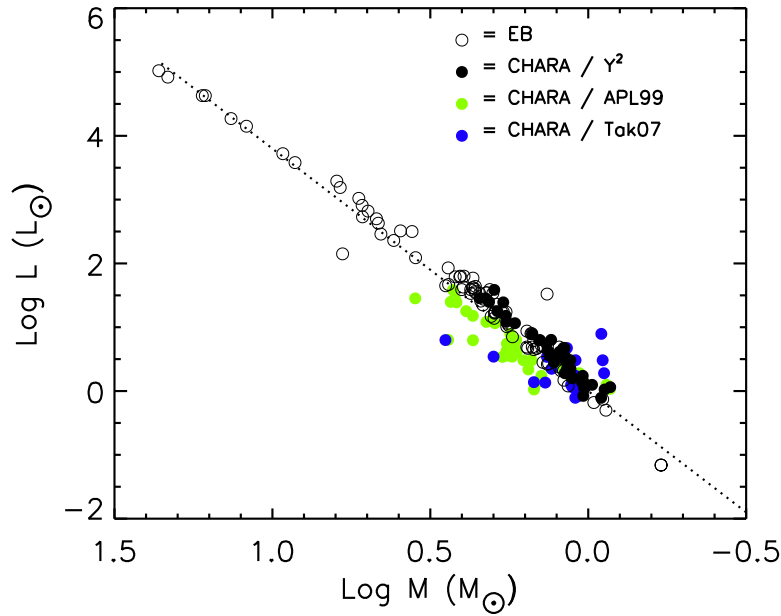}
\caption{Mass-Luminosity plot of non-evolved eclipsing binary (EB) stars in \citet{Andersen1991A&ARv...3...91A} (open circles).  Also shown are data from a large sample of main sequence solar-type stars measured with interferometry \citet{Boyajian2009PhDT........24B} where masses are found from fitting the directly determined temperatures and luminosities to the Yonsei-Yale (Y$^2$) isochrones.  Masses for these same data are solved for when applying the interferometrically measured radii with spectroscopic $\log g$ values from \citet{AllendePrieto1999A&A...352..555A} and \citet{Takeda2007PASJ...59..335T}, shown as green and blue filled points, respectively.}
\end{figure}

Ages and masses of these nearby, single, stars measured with interferometry can be found by fitting Y$^2$ isochrones \citep{Yi2001ApJS..136..417Y, Kim2002ApJS..143..499K, Demarque2004ApJS..155..667D} to the directly measured temperatures and luminosities, and these results compare very well with results from eclipsing binaries in \citet{Andersen1991A&ARv...3...91A} (see Figure~\ref{fig:M_VS_L}).  Alternatively, the masses of this same sample of stars are derived by combining spectroscopic surface gravity measurements $\log g$ with the measured interferometric radii (see Figure~\ref{fig:M_VS_L}).  However, it is apparent here that the spectroscopic $\log g$ measurements tend to be overestimated, leading to over-predicted masses (and younger ages), and is thought to be a direct consequence of the spectroscopic temperature being over-estimated.

A solid foundation to build empirical relations to stellar temperature and radius are beginning to emerge with the sensitivity of current interferometers.  Temperature scales for giants and supergiants are explored in \citet{vanBelle1999AJ....117..521V, vanBelle2009MNRAS.394.1925V} and it is shown that we have now reached accuracy of $\sim 2.5$\% level, only to be limited by the distances to these objects.  Empirical relations to the temperatures and radii of main sequence stars are reviewed from compiling the data available in \citet{vanBelle2009ApJ...694.1085V} and \citet{Boyajian2009PhDT........24B}.  Additionally we introduce new results from a large interferometric survey of late-type dwarfs at the CHARA Array \citetext{Boyajian et al., in preparation} and exoplanet host stars \citetext{von Braun et al., in preparation}, which double the number of published values to-date for these types of stars.  Statistically, there is no difference between the luminosity class of an object and the $T_{\rm EFF}$ versus ($V-K$) or $T_{\rm EFF}$ versus spectral type relations until the main sequuence stars are introduced, showing much steeper slopes, and somewhat of a dis-continuity at spectral type $\sim$~M1 ($V-K \sim 4$).  These emperical solution for a muliti-parameter $T_{\rm EFF}$:Color:Metallicity relation based on these data has now finally reached the 1\% level.

\begin{figure}[!ht]
\label{fig:M_VS_R}
\epsscale{10}
\begin{center}
\includegraphics[scale=0.5]{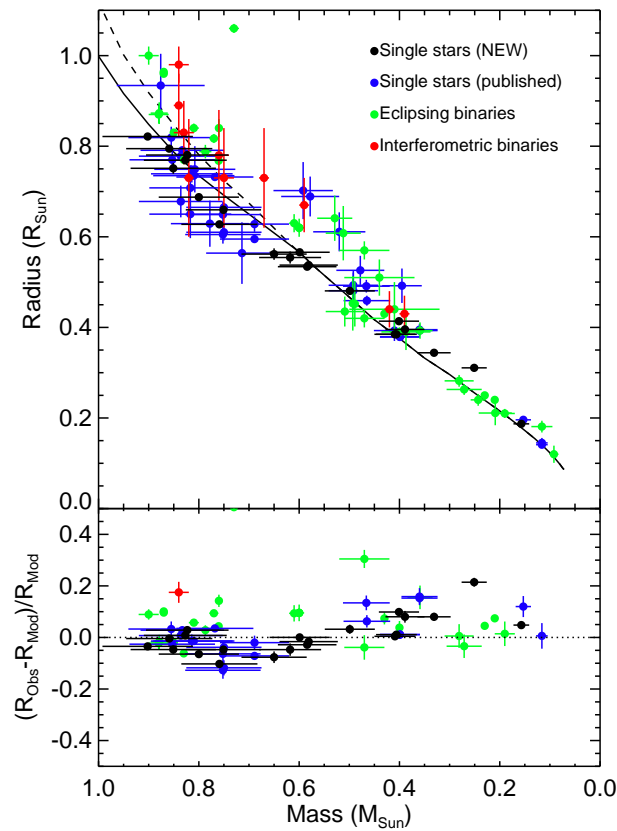}
\end{center}
\caption{{\it TOP} The Mass-Radius relationship for K and M stars. The solid black line is a 5 Gyr isochrone from the BCAH98 models \citep{Baraffe1998A&A...337..403B}. For stars with mass $> 0.6$~M$_{\odot}$, the dashed line indicates $L_{mix} = H_{p}$ and the solid line  indicates $L_{mix} = 1.9 H_{p}$.  {\it BOTTOM} Deviation in radius versus mass for stars with radii measurements better than 5\%.  Masses for single stars are derived from the $K$-band mass-luminosity relation from \citet{Delfosse2000A&A...364..217D}, and assume a 10\% error.}
\end{figure}

We conclude this discussion with a look into the of the disagreement with stellar radii as predicted by models compared to observations.  We include the (currently unpublished) work mentioned above from \citetext{Boyajian et al., in preparation} and \citetext{von Braun et al., in preparation} along with all other interferometric observations of K-M type dwarfs presented in
\citep{Lane2001ApJ...551L..81L,Segransan2003A&A...397L...5S,Boyajian2008ApJ...683..424B,Kervella2008A&A...488..667K,Demory2009A&A...505..205D,vanBelle2009ApJ...694.1085V}.  Figure~\ref{fig:M_VS_R} shows the mass-radius relationship for binary and single K and M dwarfs.  We see that for both single and binary stars, models begin to underpredict radii by an average of $\sim 10$\% (N.B. the bottom panel restricts the data plotted to radii measurements better than 5\%).  We also notice a difference in mixing lengths for stars $> 0.6 M_{\odot}$, where the single stars are modeled better with larger mixing lengths.

\subsection{Diameters and Modeling - Jason Aufdenberg}

Historically speaking, the characterizations of stellar limb darkening
have only been done for a single object - our Sun.  The Sun is the
only star for which angular resolution has been sufficiently fine to
examine limb darkening laws predicted by stellar atmosphere theory.
However, using our Sun as a guide is an imperfect approach, since this
object does not represent the full range of possibilities for stellar
atmospheric structure.

Recently interferometric observations of $\alpha$ Cen B have been
compared against the predictions of stellar limb darkening obtained
from both 1D ATLAS atmospheric models and 3D radiative hydrodynamic
(RHD) simulations \citet{Bigot2006A&A...446..635B}.  Significantly,
since convection is fundamentally a three dimensional process, the 1D
models have an inherent weakness in proper treatment of this
phenomenon.  It is no surprise that there are deviations in the 1D and
3D models, with the latter showing less limb darkening - resulting in
diameters that are 0.1\% smaller in the near-IR, and up to 1.5\%
smaller in the optical.

Further observational evidence that we are departing the era where 1D
models are sufficient is seen in the data on Procyon presented in
\citet{Aufdenberg2005ApJ...633..424A}.  Comparison of models to the
measured stellar diameters of Procyon at both the optical and near-IR
wavelengths shows a good match to 3D model predictions as well as 1D
models which include an overshooting correction.  Furthermore, the
nature of 3D models allow for a range of temperature structures,
consistent with granulation structure (like that which is known to
appear on the solar surface) and its multi-spectral nature, required
to match Procyon's spectral energy distribution.  This also appears to
be the case for 3D models of late-type giants where 3-D models show
significant flux deviations ($\sim$5-20\%) in the visible and
shortwards, along with notable deviations in the near-infrared
($\sim$5\%) \citep{Kucinskas2009MmSAI..80..723K} relative to 1D
models.  Precision angular diameter measurements are needed in the
optical, for example on $\alpha$ Cen B (K1 V), where the 3D convection
models now applied to our Sun can be further tested.

Early closure phase imaging experiments have shown that giant and
supergiant stars have mottled, inhomogenous surfaces
\citep{Tuthill1997MNRAS.285..529T}. Recent comparisons of
interferometric data against 1D models have show increasing
differences with increasing spatial frequency data
\citep{Haubois2009A&A...508..923H}, indicative of poor treatment of
convection.  In contrast, 3D RHD models show much better agreement
when confronted by similar data sets
\citep{Chiavassa2009A&A...506.1351C,Chiavassa2010A&A...515A..12C}.
Extending these 3D models from gray opacities to multi-wavelength
radiative transfer is a necessary next step. The current use of gray
opacities means the thermal gradient is likely too shallow, and
neglect of radiation pressure means the predicted atmosphere is
probably too compact.  Multi-wavelength angular diameters of objects
like $\alpha$ Ori in the near- and mid-IR
\citep{Perrin2004A&A...418..675P} indicate a strong dependence on
diameter with wavelength, a dependence only exacerbated at higher
spectral resolution
\citep{Wittkowski2004A&A...413..711W,Wittkowski2006A&A...460..855W}.

\begin{figure}[!ht]
\label{fig:procyon}
\epsscale{10}
\plotone{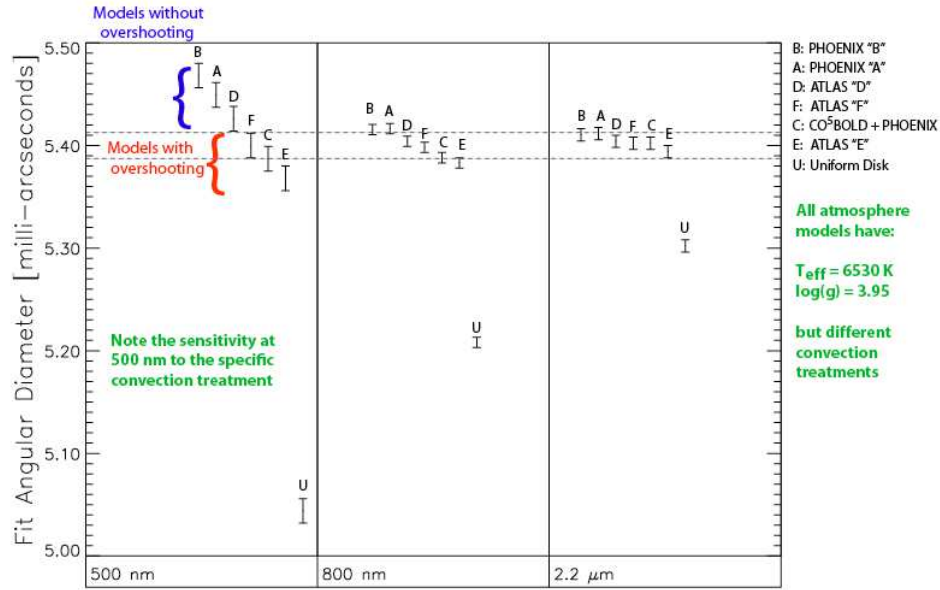}
\caption{Comparison of the best-fit angular diameters of Procyon at
500 nm, 800nm and 2.2 microns for seven atmosphere models with
different convection treatments.  Based on a figure from
\citet{Aufdenberg2005ApJ...633..424A}}
\end{figure}

\section{Orbits and Masses - Christian Hummel}

Arguably the most fundamental parameter for a star is its mass, as this sets the timescale for evolution and determines the star's ultimate fate.  The high resolution capabilities of optical interferometry have the potential to greatly increase the number and types of stars for which we have dynamical mass estimates, and greatly improve overall mass estimates.  The traditional technique of using eclipsing binaries to provide these measurements limits studies to biased samples where a history of mass exchange is more likely - high angular resolution is needed to have the freedom to observe ``typical'' binaries

Although an early example of interferometric observations of a binary can be found as early as 40 years ago \citep{HanburyBrown1970MNRAS.148..103H}, routine modern observations began roughly two decades later with small-aperture interferometers in the visual and NIR going after easy targets to show the power of milli-arcsec resolution, with a limiting magnitude about V=5 \citep[e.g.][]{Armstrong1992AJ....104.2217A}.
After many successful orbit determinations on systems that were `forgiving' from the standpoint of brightness and/or orbital parameters, the emphasis is now being put on observing astrophysically more challenging and interesting targets.
Simply put, interferometry of these objects is now a mature technique offered in service mode at Keck and VLTI.

\begin{figure}\label{fig_kraus}
\plottwo{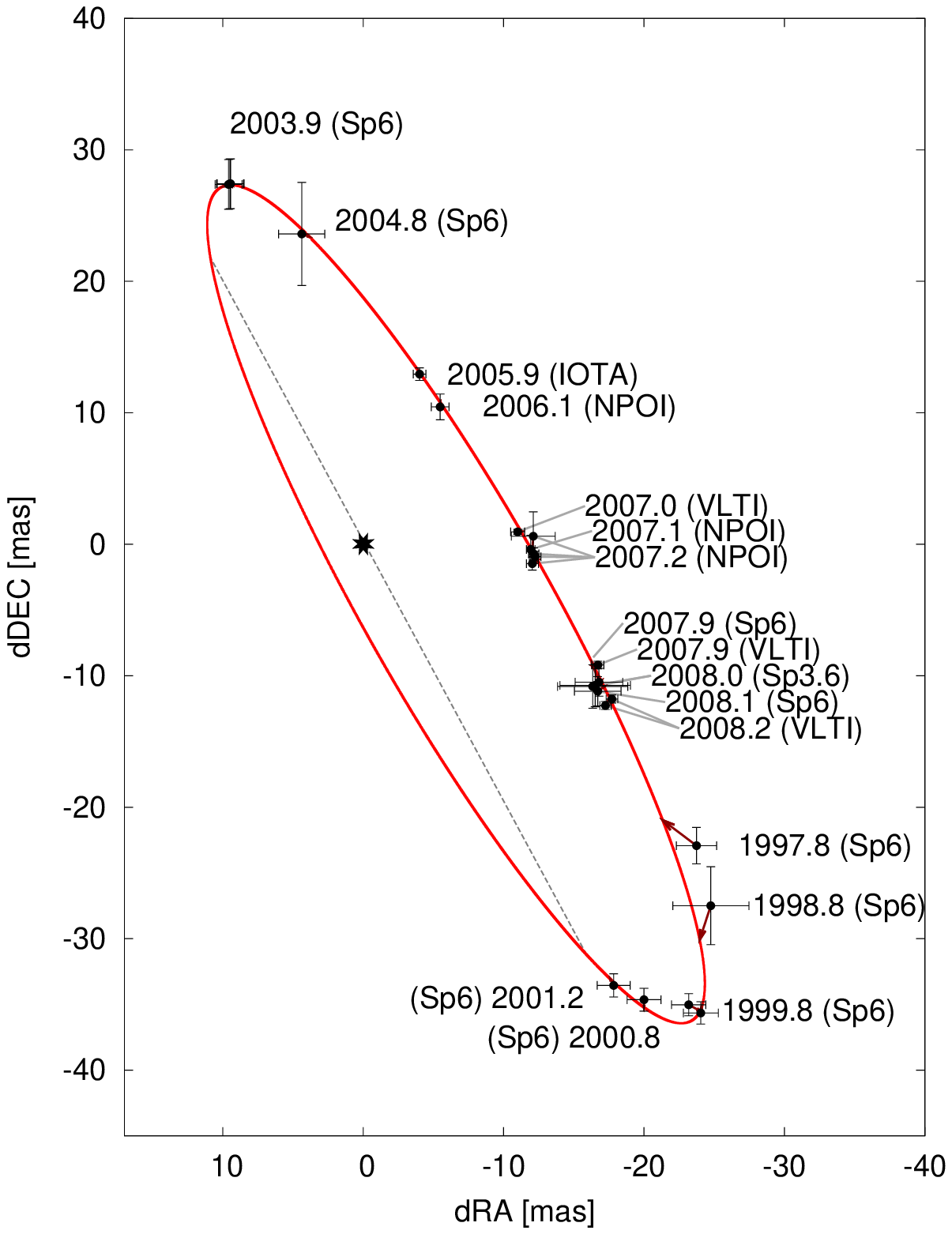}{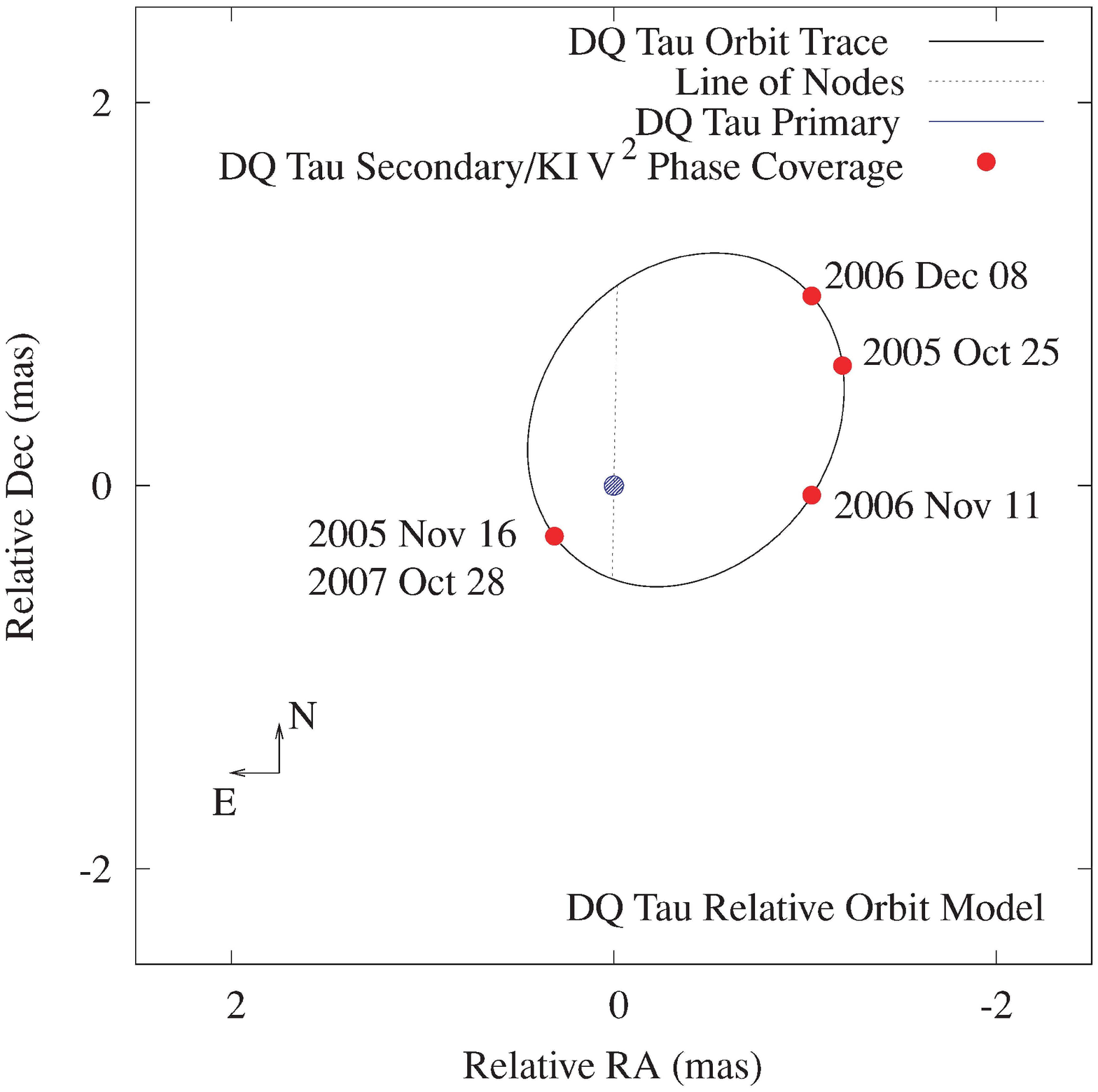}
\caption{Observations of binaries being done with modern interferometers: (Left) the orbit of $\theta^1$ Ori C, from Figure 8 from \citet{Kraus2009A&A...497..195K}, using data from IOTA/NPOI/VLTI; (right) the orbit of DQ Tau from Figure 3 of \citet{Boden2009ApJ...696L.111B} with KI data.}
\end{figure}

Dynamical masses of young binaries from interferometry
\citep{Boden2005ApJ...627..464B,Schaefer2008AJ....135.1659S}
are an area where only interferometry allows the study of non-eclipsing systems: mass estimates at this early age are especially important because of the poorly understood input physics (e.g.  convection, opacities) of pre-main sequence stars.  With precise distance estimates \citep[some from FGS interferometry on board HST, e.g.][]{Benedict2006AJ....132.2206B} relative orbits will provide dynamical masses of stars in rarer evolutionary states, such as those transitioning to the giant phases \citep{Boden2005ApJ...635..442B} and traversing the Hertzsprung Gap \citep{Boden2006ApJ...644.1193B}.  Additionally, only just recently have the prospects of high contrast imaging via non-redundant aperture masking on large-diameter telescopes been realized.  This work can provide dynamical masses for sub-stellar objects \citep{Ireland2008ApJ...678..463I}, a mass range where evolutionary models are very poorly constrained due to the age/temperature/mass degeneracies.
As an example of the current state of the art, the results found in \citet{Kraus2009A&A...497..195K} and \citet{Boden2009ApJ...696L.111B}
are seen in Figure \ref{fig_kraus}.


References on binary star work that are useful include the general review of optical interferometry in astronomy by \citet{Monnier2003RPPh...66..789M} and the recent examination of accurate masses and radii of normal stars by\citet{Torres2010A&ARv..18...67T}.  Online, the web sites for ``Optical Long Baseline Interferometry News''\footnote{http://olbin.jpl.nasa.gov/index.html} and ``Database of Publications in Stellar Interferometry''\footnote{http://apps.jmmc.fr/bibdb/} are both useful.

\section{Recent Observational Results and Advances}

\subsection{CHARA MIRC - Ettore Pedretti}

The Georgia State University (GSU) Center for High Angular Resolution Astronomy (CHARA) Array is a six-element optical interferometer \citep{tenBrummelaar2005ApJ...628..453T} that is beginning to produce true dilute aperture images.  With the Michigan Infrared Combiner (MIRC) instrument \citep{Monnier2006SPIE.6268E..55M}, initial 4-way imaging has begun to produce remarkable results.  It is further interesting to note that the VLBA has a physical size of roughly 10,000km (Hawaii to Puerto Rico), while its optical analog in the CHARA Array spans only 330m - and yet it has comparable angular resolution, given the shorter wavelength of operation.  The capability to directly image of the surfaces of hotter stars such as Altair \citep{Monnier2007Sci...317..342M} and $\alpha$ Oph \& $\alpha$ Cep \citep{Zhao2009ApJ...701..209Z} is now being directed towards cooler objects such as the cloud occulting the $\epsilon$ Aur \citep{Kloppenborg2010Natur.464..870K} or $\zeta$ And.  The latter object is of particular interest given availability of Doppler imaging data for comparison \citep{Kovari2007A&A...463.1071K}.
An upgrade to full 6-telescope operation will further enhance the `snapshot' capabilities of the system \citep{Monnier2010SPIE.7734E..13M}.

\subsection{The `A-List' - Ellyn Baines}

In addition to their many other peculiarities, the ages of A-type stars are poorly known.  While these objects are on the main sequence, they evolve in $T_{EFF}$ and in $L/R$, which affords an opportunity to establish their ages through interferometric means.  By obtaining their angular sizes, and deriving values for $R$ and $T_{EFF}$, observational data can be compared to models on an H-R diagram, which will then indicate their ages and masses.  Such investigations are particularly timely given the recent discovery of potential planetary objects orbiting HR8799 \citep{Marois2008Sci...322.1348M,Marois2010arXiv1011.4918M}.

For this particular star, its estimated age is 30-160Myr - based on a evidence that only poorly constrains that parameter: the object's galactic space motion, its placement on a color-magnitude diagram, and `typical' ages for $\lambda$ Boo or $\gamma$ Dor stars \citep[][and references therein]{Marois2008Sci...322.1348M}.  The current age estimate of $\sim$60Myr for HR8799 is presented with considerable uncertainty.

\citet{Marois2008Sci...322.1348M} indicate that an age of $>300$Myr would be necessary for all companion objects to be brown dwarfs. An initial investigation of this object using the CHARA Array has preliminary results that could be consistent with this object being well in excess of this age; this result is consistent with the asteroseismic finding of \citet{Moya2010MNRAS.405L..81M}.

\subsection{Young Star Sizes - Russel White}

Using temperature, photometry, and distance measurements of young stars in the nearest and youngest moving groups ($\beta$ Pictoris and AB Dor), we predict the stellar radii and angular sizes of known members to identify if any can be spatially resolved with the long-baseline near-infrared interferometers operating in the northern and southern hemispheres.  The motivation is to potentially constrain pre-main sequence evolutionary models from direct radius measurements.  In the northern  hemisphere ($\delta > -20^o$), 3 stars have sizes large enough to be  spatially resolved ($\theta \gtrsim 0.4$) with the CHARA Array, which has a baseline of 331-m; this subsample includes the low mass M2 dwarf GJ 393 which is near the fully convective boundary. All 3 have now been successfully spatially resolved with a precision of of a few percent; the analysis of these results is still maturing. In the southern hemisphere ($\delta < +20^o$), 9 stars have sizes larger than this angular diameter, including the high-profile low mass debris disk star AU Mic.  However, the current longest baselines of the VLTI are not able to resolve any of these systems; opening new longer-baseline stations may alleviate this problem.

\subsection{Betelgeuse with the ISI, 2006-2009 - Vikram Ravi}

Closure phase observations of Betelgeuse ($\alpha$ Ori) at 11 $\mu$m over the four year interval 2006-2009 have been able to probe surface features and the size of this object as it evolves in time \citep{Wishnow2010SPIE.7734E...8W}.  During this interval, contributions of individual spots to the overall surface appear with a measure of time variability, possibly consistent with evolution of convection cells on the surface.  The effective surface temperature of the star appears to drift slightly in time as well, from $2750\pm100$K in 2006 to $2350\pm250$K in 2008, back up to $2650\pm75$ in 2009.  This temperature variation is found to be directly correlated with the size of the star. Ongoing operations of the ISI for 2010-2011 include a upgrade for very high spectral resolution, which can simultaneously measure spectral lines and continuum.

\section{The `Outsider Perspective' on Where Interferometry Can Contribute - \\Graham~Harper}

There are a number of areas where even a small number of `magic bullet' interferometric observations can make significant contributions.  (This list is intended to be instructive rather than comprehensive.)

{\bf \#1 Break the photospheric models}.  For the current stock of photospheric models a certain number of assumptions are inherent in their construction - assumptions which do not always hold good.  For example, outer boundary conditions, assumptions of hydrostatic equilibrium, dimensionality of model (e.g. 1D versus 3D), all can lead to weaknesses in the models - weakness which can be exposed by probing those stellar candidates that are most sensitive to such assumptions.  The best current candidates are bright giants \& supergiants, which have lower surface gravity and less convective cells; later spectral types are also better in that the influence of molecules and neutral opacity goes up.
The best observational techniques to employ here are measurements of wavelength-dependent limb darkening, taking data beyond the first null of the visibility curve.

{\bf \#2 Even more accurate stellar $\theta_{\rm LD}$}.  Specifically, measures where $\Delta \theta / \theta \leq$1-2\%.  There are a number applications: first spectroscopic absolute flux calibrations are reaching the $\sim$5\% level, and stellar chromospheric surface fluxes can be measured at this level. These reflect the radiative cooling that matches the chromospheric heating, whose origin is a source of active research. Since $F_\oplus \propto \theta^2 F_\star$, and $\Delta F_\star / F_\star \sim 2 \Delta \theta / \theta + \Delta F_\oplus / F_\oplus$, the desire is for 1-2\% errors on $\theta$ to match the flux errors.  Second, when considering ALMA observations, errors in radio brightness temperature ($T_{\rm Br}$) are related to angular size errors in a way that is directly analogous to our first case, e.g. $\Delta T_{\rm Br} / T_{\rm Br}\sim 2 \Delta \theta / \theta + \Delta F / F$.  However, when considering ALMA maps where the errors in mean electron temperature, $T_e$, are directly proportional to $T_{\rm Br}$, a 2\% error in $T_e$ in the mid-chromosphere can lead to errors in electron density of $\Delta n_e / n_e \sim $100\%.

{\bf \#3 Calibrate surface brightness relations}.
In particular, for late-type low amplitude variability K, M and C-stars.  An important question is, what is limiting the precision of existing surface brightness relations? For the foreseeable future there are always going to be stars that do not have measured angular diameters, so it is
important to understand the limitations of such indirect techniques.

{\bf \#4 Geometry and gravity darkening}.
The rotationally oblate star Altair exhibits unusual geometry and gravity darkening due to its rapid rotation \citep{vanBelle2001ApJ...559.1155V,Ohishi2004ApJ...612..463O,Monnier2007Sci...317..342M}.  Similarly, gravitational darkening may be present in eclipsing binary $\zeta$ Aurigae systems distorted near periastron \citep{Guinan1979PASP...91..343G,Eaton2008ApJ...679.1490E}.  Resolving out the orbital geometry {\it and} gravity darkening of such systems would be instructive for
stellar structure studies.

In summary, there are a number of things that are desirable to know.  Specifically, and non-exhaustively:
(1) How significant are existing differences in limb-darkening fits?
(2) For what spectral-types and luminosity classes do MOL-spheres exist. What is their distribution in the HR diagram?
(3) Can we resolve the geometry/gravity darkening in convective low gravity stars?
(4) What limits the precision of surface brightness relations?


Of particular value for addressing such questions raised by non-interferometrists, in a timely fashion,  would be some mechanism for requesting $\theta_{\rm LD}$ for ``special interest'' stars. Perhaps a simple web-based system could be established where astronomers could post their needs and potential collaborations identified.

\section{Opportunities for Observers - Steve Ridgway}
Currently there are a number of facilities world-wide for which optical interferometric observations are possible.
There are two facility-class installations (Keck Interferomeer, VLTI) with open calls, and a third instrument with some open time (CHARA).   Additionally, there are three that are primarily accessible through collaborations with members (CHARA, NPOI, SUSI).
Additionally, a substantial body of unpublished archival data from the now-closed Palomar Testbed Interferometer is available online.



\section{Conclusion}

Optical interferometry is a challenging technique that rewards those who take up that challenge with observational data unobtainable via any other approach.  In particular, contributions of the technique in the area of fundamental parameters of cool stars are substantial, highlighting areas of concern and interest for theorists.  Furthermore, the facilities that provide access to this technique have matured substantially over the last 10 years, making it more accessible to all manners of astronomer.

\acknowledgements The splinter organizers would like to thank the splinter organizing committee (Rachel Akeson,
Jason Aufdenberg,
Andrew Boden,
Tabetha Boyajian,
Michelle Creech-Eakman,
Christian Hummel,
Stephen Ridgway, Peter Tuthill, GvB) for their contributions to the development of the splinter program, and to the organizers of CS16 for providing the splinter opportunity.

\bibliography{cs16vB}

\end{document}